# Artificial Intelligence: the global landscape of ethics guidelines

Anna Jobin [a], Marcello Ienca [a], Effy Vayena [a]*

[a] Health Ethics & Policy Lab, ETH Zurich, 8092 Zurich, Switzerland

* Corresponding author: effy.vayena@hest.ethz.ch

## ABSTRACT

In the last five years, private companies, research institutions as well as public sector organisations have issued principles and guidelines for ethical AI, yet there is debate about both what constitutes "ethical AI" and which ethical requirements, technical standards and best practices are needed for its realization. To investigate whether a global agreement on these questions is emerging, we mapped and analyzed the current corpus of principles and guidelines on ethical AI. Our results reveal a global convergence emerging around five ethical principles (transparency, justice and fairness, non-maleficence, responsibility and privacy), with substantive divergence in relation to how these principles are interpreted; why they are deemed important; what issue, domain or actors they pertain to; and how they should be implemented. Our findings highlight the importance of integrating guideline-development efforts with substantive ethical analysis and adequate implementation strategies.

# MAIN ARTICLE

## Introduction

Artificial Intelligence (AI), or the theory and development of computer systems able to perform tasks normally requiring human intelligence, is widely heralded as an ongoing "revolution" transforming science and society altogether[1,2]. While approaches to AI such as machine learning, deep learning and artificial neural networks are reshaping data processing and analysis[3], autonomous and semi-autonomous systems are being increasingly used in a variety of sectors including healthcare, transportation and the production chain[4]. In light of its powerful transformative force and profound impact across various societal domains, AI has sparked ample debate about the principles and values that should guide its development and use[5,6]. Fears that AI might jeopardize jobs for human workers[7], be misused by malevolent actors[8], elude accountability or inadvertently disseminate bias and thereby undermine fairness[9] have been at the forefront of the recent scientific literature and media coverage. Several studies have discussed the topic of ethical AI[10–13], notably in meta-assessments[14–16] or in relation to systemic risks[17,18] and unintended negative consequences like algorithmic bias or discrimination[19–21].

National and international organisations have responded to these societal fears by developing *ad hoc* expert committees on AI, often commissioned with the drafting of policy documents. These include the High-Level Expert Group on Artificial Intelligence appointed by the European Commission, the expert group on AI in Society of the Organisation for Economic Co-operation and Development (OECD), the Advisory Council on the Ethical Use of Artificial Intelligence and Data in Singapore, and the select committee on Artificial Intelligence of the United Kingdom (UK) House of Lords. As part of their institutional appointments, these committees have produced or are reportedly producing reports and guidance documents on AI. Similar efforts are taking place in the private sector, especially among corporations who rely on AI for their business. In 2018 alone, companies such as Google and SAP have publicly released AI guidelines and principles. Declarations and recommendations have also been issued by professional associations and non-profit organisations such as the Association of Computing Machinery (ACM), Access Now and Amnesty International. The intense efforts of such a diverse set of stakeholders in issuing AI principles and policies demonstrate not only the need for ethical guidance, but also the strong interest of these stakeholders to shape the ethics of AI in ways that meet their respective priorities[16]. Notably, the private sector's involvement in the AI-ethics arena has been called into question for potentially using such high-level soft-policy as a portmanteau to either render a social problem technical[16] or to eschew regulation altogether[22]. Beyond the composition of the groups that have produced ethical guidance on AI, the content of this guidance itself is of interest. Are these various groups converging on what ethical AI should be, and the ethical principles that will determine the development of AI? If they diverge, what are these differences and can they be reconciled?

To answer these questions, we conducted a scoping review of the existing corpus of guidelines on ethical AI. Our analysis aims at mapping the global landscape of existing principles and guidelines for ethical AI and thereby determining whether a global convergence is emerging regarding both the principles for ethical AI and the requirements for its realization. This analysis will inform scientists, research institutions, funding agencies, governmental and inter-governmental organisations and other relevant stakeholders involved in the advancement of ethically responsible innovation in AI.

## Results

Our search identified 84 documents containing ethical principles or guidelines for AI (cf. Table 1). Data reveal a significant increase over time in the number of publications, with 88% having been released after 2016 (cf. SI Table S1). Data breakdown by type and geographic location of issuing organisation (cf. SI Table S1) shows that most documents were produced by private companies (n=19; 22.6%) and governmental agencies respectively (n=18; 21.4%), followed by academic and research institutions (n=9; 10.7%), inter-governmental or supra-national organisations (n=8; 9.5%), non-profit organisations and professional associations/scientific societies (n=7 each; 8.3% each), private sector alliances (n=4; 4.8%), research alliances (n=1; 1.2%), science foundations (n=1; 1.2%), federations of worker unions (n=1; 1.2%) and political parties (n=1; 1.2%). Four documents were issued by initiatives belonging to more than one of the above categories and four more could not be classified at all (4.8% each).

*Table 1- Ethical guidelines for AI by country of issuer*

| **Name of Document/Website** | **Issuer** | **Country of issuer** |
|---|---|---|
| Artificial Intelligence. Australia's Ethics Framework. A discussion Paper | Department of Industry Innovation and Science | Australia |
| Montréal Declaration: Responsible AI | Université de Montréal | Canada |
| Work in the age of artificial intelligence. Four perspectives on the economy, employment, skills and ethics | Ministry of Economic Affairs and Employment | Finland |
| Tieto's AI ethics guidelines | Tieto | Finland |
| Commitments and principles | OP Group | Finland |
| How can humans keep the upper hand? Report on the ethical matters raised by AI algorithms | French Data Protection Authority (CNIL) | France |
| For a meaningful Artificial Intelligence. Towards a French and European strategy | Mission Villani | France |
| Ethique de la recherche en robotique | CERNA (Allistene) | France |
| AI Guidelines | Deutsche Telekom | Germany |
| SAP's guiding principles for artificial intelligence | SAP | Germany |
| Automated and Connected Driving: Report | Federal Ministry of Transport and Digital Infrastructure, Ethics Commission | Germany |
| Ethics Policy | Icelandic Institute for Intelligent Machines (IIIM) | Iceland |
| Discussion Paper: National Strategy for Artificial Intelligence | National Institution for Transforming India (Niti Aayog) | India |
| L'intelligenzia artificiale al servizio del cittadino | Agenzia per l'Italia Digitale (AGID) | Italy |
| The Japanese Society for Artificial Intelligence Ethical Guidelines | Japanese Society for Artificial Intelligence | Japan |
| Report on Artificial Intelligence and Human Society (Unofficial translation) | Advisory Board on Artificial Intelligence and Human Society (initiative of the Minister of State for Science and Technology Policy) | Japan |
| Draft AI R&D Guidelines for International Discussions | Institute for Information and Communications Policy (IICP), The Conference toward AI Network Society | Japan |
| Sony Group AI Ethics Guidelines | SONY | Japan |
| Human Rights in the Robot Age Report | The Rathenau Institute | Netherlands |
| Dutch Artificial Intelligence Manifesto | Special Interest Group on Artificial Intelligence (SIGAI), ICT Platform Netherlands (IPN) | Netherlands |
| Artificial intelligence and privacy | The Norwegian Data Protection Authority | Norway |
| Discussion Paper on Artificial Intelligence (AI) and Personal Data - Fostering Responsible Development and Adoption of AI | Personal Data Protection Commission Singapore | Singapore |
| Mid- to Long-Term Master Plan in Preparation for the Intelligent Information Society | Government of the Republic of Korea | South Korea |
| AI Principles of Telefónica | Telefonica | Spain |
| AI Principles & Ethics | Smart Dubai | UAE |
| Principles of robotics | Engineering and Physical Sciences Research Council UK (EPSRC) | UK |
| The Ethics of Code: Developing AI for Business with Five Core Principles | Sage | UK |
| Big data, artificial intelligence, machine learning and data protection | Information Commissioner's Office | UK |
| DeepMind Ethics & Society Principles | DeepMind Ethics & Society | UK |
| Business Ethics and Artificial Intelligence | Institute of Business Ethics | UK |
| AI in the UK: ready, willing and able? | UK House of Lords, Select Committee on Artificial Intelligence | UK |
| Artificial Intelligence (AI) in Health | Royal College of Physicians | UK |
| Initial code of conduct for data-driven health and care technology | UK Department of Health & Social Care | UK |
| Ethics Framework - Responsible AI | Machine Intelligence Garage Ethics Committee | UK |
| The responsible AI framework | PriceWaterhouseCoopers UK | UK |
| Responsible AI and robotics. An ethical framework. | Accenture UK | UK |
| Machine learning: the power and promise of computers that learn by example | The Royal Society | UK |
| Ethical, social, and political challenges of Artificial Intelligence in Health | Future Advocacy | UK |
| Unified Ethical Frame for Big Data Analysis. IAF Big Data Ethics Initiative, Part A | The Information Accountability Foundation | UK |
| The AI Now Report. The Social and Economic Implications of Artificial Intelligence Technologies in the Near-Term | AI Now Institute | USA |
| Statement on Algorithmic Transparency and Accountability | Association for Computing Machinery (ACM) | USA |
| AI Principles | Future of Life Institute | USA |
| AI - Our approach | Microsoft | USA |
| Artificial Intelligence. The Public Policy Opportunity | Intel Corporation | USA |
| IBM's Principles for Trust and Transparency | IBM | USA |
| OpenAI Charter | OpenAI | USA |
| Our principles | Google | USA |
| Policy Recommendations on Augmented Intelligence in Health Care H-480.940 | American Medical Association (AMA) | USA |
| Everyday Ethics for Artificial Intelligence. A practical guide for designers & developers | IBM | USA |
| Governing Artificial Intelligence. Upholding Human Rights & Dignity | Data & Society | USA |
| Intel's AI Privacy Policy White Paper. Protecting individuals' privacy and data in the artificial intelligence world | Intel Corporation | USA |
| Introducing Unity's Guiding Principles for Ethical AI – Unity Blog | Unity Technologies | USA |
| Digital Decisions | Center for Democracy & Technology | USA |
| Science, Law and Society (SLS) Initiative | The Future Society | USA |
| AI Now 2018 Report | AI Now Institute | USA |
| Responsible bots: 10 guidelines for developers of conversational AI | Microsoft | USA |
| Preparing for the future of Artificial Intelligence | Executive Office of the President; National Science and Technology Council; Committee on Technology | USA |
| The National Artificial Intelligence Research and Development Strategic Plan | National Science and Technology Council; Networking and Information Technology Research and Development Subcommittee | USA |
| AI Now 2017 Report | AI Now Institute | USA |
| Position on Robotics and Artificial Intelligence | The Greens (Green Working Group Robots) | EU |
| Report with recommendations to the Commission on Civil Law Rules on Robotics | European Parliament | EU |
| Ethics Guidelines for Trustworthy AI | High-Level Expert Group on Artificial Intelligence | EU |
| AI4People—An Ethical Framework for a Good AI Society: Opportunities, Risks, Principles, and Recommendations | AI4People | EU |
| European ethical Charter on the use of Artificial Intelligence in judicial systems and their environment | Concil of Europe: European Commission for the efficiency of Justice (CEPEJ) | EU |
| Statement on Artificial Intelligence, Robotics and 'Autonomous' Systems | European Commission, European Group on Ethics in Science and New Technologies | EU |
| Artificial Intelligence and Machine Learning: Policy Paper | Internet Society | international |
| Report of COMEST on Robotics Ethics | COMEST/UNESCO | international |
| Ethical Principles for Artificial Intelligence and Data Analytics | Software & Information Industry Association (SIIA), Public Policy Division | international |
| ITI AI Policy Principles | Information Technology Industry Council (ITI) | international |
| Ethically Aligned Design. A Vision for Prioritizing Human Well-being with Autonomous and Intelligent Systems, version 2 | Institute of Electrical and Electronics Engineers (IEEE), The IEEE Global Initiative on Ethics of Autonomous and Intelligent Systems | international |
| Top 10 Principles for Ethical Artificial Intelligence | UNI Global Union | international |
| The Malicious Use of Artificial Intelligence: Forecasting, Prevention, and Mitigation | Future of Humanity Institute; University of Oxford; Centre for the Study of Existential Risk; University of Cambridge; Center for a New American Security; Electronic Frontier Foundation; OpenAI | international |
| White Paper: How to Prevent Discriminatory Outcomes in Machine Learning | WEF, Global Future Council on Human Rights 2016-2018 | international |
| Privacy and Freedom of Expression In the Age of Artificial Intelligence | Privacy International & Article 19 | international |
| The Toronto Declaration: Protecting the right to equality and non-discrimination in machine learning systems | Access Now ; Amnesty International | international |
| Charlevoix Common Vision for the Future of Artificial Intelligence | Leaders of the G7 | international |
| Artificial Intelligence: open questions about gender inclusion | W20 | international |
| Declaration on ethics and data protection in Artificial Intelligence | ICDPPC | international |
| Universal Guidelines for Artificial Intelligence | The Public Voice | international |
| Ethics of AI in Radiology: European and North American Multisociety Statement | American College of Radiology; European Society of Radiology; Radiology Society of North America; Society for Imaging Informatics in Medicine; European Society of Medical Imaging Informatics; Canadian Association of Radiologists; American Association of Physicists in Medicine | international |
| Ethically Aligned Design: A Vision for Prioritizing Human Well-being with Autonomous and Intelligent Systems, First Edition (EAD1e) | Institute of Electrical and Electronics Engineers (IEEE), The IEEE Global Initiative on Ethics of Autonomous and Intelligent Systems | international |
| Tenets | Partnership on AI | n.a. |
| Principles for Accountable Algorithms and a Social Impact Statement for Algorithms | Fairness, Accountability, and Transparency in Machine Learning (FATML) | n.a. |
| 10 Principles of responsible AI | Women leading in AI | n.a. |

In terms of geographic distribution, data show a significant representation of more economically developed countries (MEDC), with the USA (n=20; 23.8%) and the UK (n=14; 16.7%) together accounting for more than a third of all ethical AI principles, followed by Japan (n=4; 4.8%), Germany, France, and Finland (each n=3; 3.6% each). The cumulative number of sources from the European Union, comprising both documents issued by EU institutions (n=6) and documents issued within each member state (13 in total), accounts for 19 documents overall. African and South-American countries are not represented independently from international or supra-national organisations (cf. Figure 1).

*Figure 1- Geographic distribution of issuers of ethical AI guidelines by number of documents released*

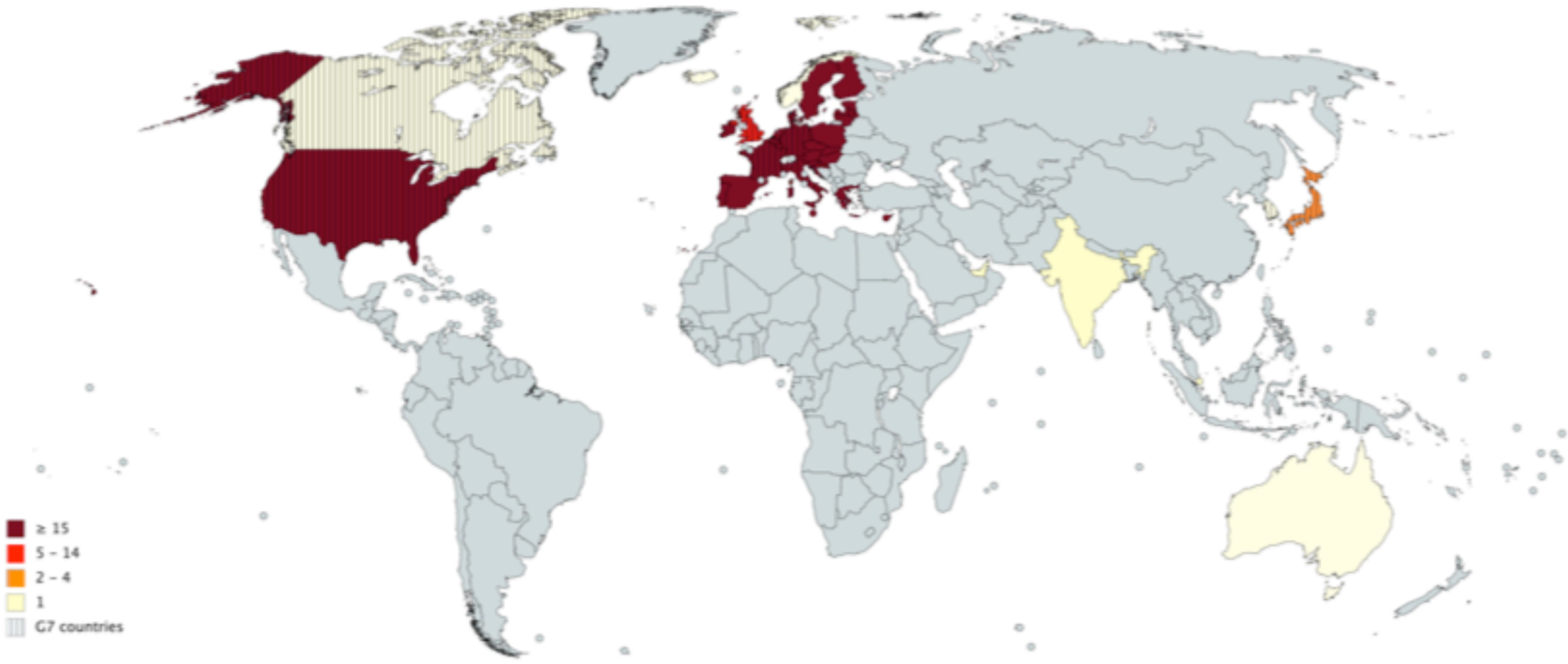

Figure 1: Geographic distribution of issuers of ethical AI guidelines by number of documents released. Most ethics guidelines are released in the United States (n=20) and within the European Union (19), followed by the United Kingdom (14) and Japan (4). Canada, Iceland, Norway, the United Arab Emirates, India, Singapore, South Korea, Australia are represented with 1 document each. Having endorsed a distinct G7 statement, member states of the G7 countries are highlighted separately. Map created using mapchart.net.

Data breakdown by target audience indicates that most principles and guidelines are addressed to multiple stakeholder groups (n=27; 32.1%). Another significant portion of the documents is self-directed, as they are addressed to a category of stakeholders within the sphere of activity of the issuer such as the members of the issuing organisation or the issuing company's employees (n=24; 28.6%). Finally, some documents target the public sector (n=10; 11.9%), the private sector (n=5; 6.0%), or other specific stakeholders beyond members of the issuing organisation, namely developers or designers (n=3; 3.6%), 'organisations' (n=1; 1.2%) and researchers (n=1; 1.2%). 13 sources (15.5%) do not specify their target audience (cf. SI Table S1).

Eleven overarching ethical values and principles have emerged from our content analysis. These are, by frequency of the number of sources in which they were featured: transparency, justice and fairness, non-maleficence, responsibility, privacy, beneficence, freedom and autonomy, trust, dignity, sustainability, and solidarity (cf. Table 2).

*Table 2 – Ethical principles identified in existing AI guidelines*

| Ethical principle | Number of documents | Included codes |
|---|---|---|
| Transparency | 73/84 | Transparency, explainability, explicability, understandability, interpretability, communication, disclosure, showing |
| Justice & fairness | 68/84 | Justice, fairness, consistency, inclusion, equality, equity, (non-)bias, (non-)discrimination, diversity, plurality, accessibility, reversibility, remedy, redress, challenge, access and distribution |
| Non-maleficence | 60/84 | Non-maleficence, security, safety, harm, protection, precaution, prevention, integrity (bodily or mental), non-subversion |
| Responsibility | 60/84 | Responsibility, accountability, liability, acting with integrity |
| Privacy | 47/84 | Privacy, personal or private information |
| Beneficence | 41/84 | Benefits, beneficence, well-being, peace, social good, common good |
| Freedom & autonomy | 34/84 | Freedom, autonomy, consent, choice, self-determination, liberty, empowerment |
| Trust | 28/84 | Trust |
| Sustainability | 14/84 | Sustainability, environment (nature), energy, resources (energy) |
| Dignity | 13/84 | Dignity |
| Solidarity | 6/84 | Solidarity, social security, cohesion |

No single ethical principle appeared to be common to the entire corpus of documents, although there is an emerging convergence around the following principles: transparency, justice and fairness, non-maleficence, responsibility, and privacy. These principles are referenced in more than half of all the sources. Nonetheless, further thematic analysis reveals significant semantic and conceptual divergences in both how the eleven ethical principles are interpreted and the specific recommendations or areas of concern derived from each. A detailed thematic evaluation is presented in the following.

***Transparency***

Featured in 73/84 sources, transparency is the most prevalent principle in the current literature. Thematic analysis reveals significant variation in relation to the interpretation, justification, domain of application, and mode of achievement. References to transparency comprise efforts to increase explainability, interpretability or other acts of communication and disclosure (cf. Table 2). Principal domains of application include data use[23–26], human-AI interaction[23,27–35], automated decisions[26,36–46], and the purpose of data use or application of AI systems[24,27,47–51]. Primarily, transparency is presented as a way to minimize harm and improve AI[36–38,44,45,49,52–55], though some sources underline its benefit for legal reasons[37,45,46,49,50,52] or to foster trust[23,24,29,33,36,37,48,51,52,56–58]. A few sources also link transparency to dialogue, participation, and the principles of democracy[30,41,49,50,52,59].

To achieve greater transparency, many sources suggest increased disclosure of information by those developing or deploying AI systems[36,51,60,61], although specifications regarding *what* should be communicated vary greatly: use of AI[45], source code[31,52,62], data use[35,47,50,58], evidence base for AI use[57], limitations[25,33,47,51,58,60,63], laws[62,64], responsibility for AI[40], investments in AI[44,65] and possible impact[66]. The provision of explanations 'in non-technical terms'[26] or auditable by humans[37,60] is encouraged. Whereas audits and auditability[28,39,44,45,50,59,61,62,67,68] are mainly proposed by data protection offices and NPOs, it is mostly the private sector that suggests technical solutions[27,30,52,59,69,70]. Alternative measures focus on oversight[45,47,48,55,62], interaction and mediation with stakeholders and the public[24,32,36,51,61,71] and the facilitation of whistleblowing[36,60].

***Justice, fairness, and equity***

Justice is mainly expressed in terms of fairness[23,25,27–29,48,50,58,60,66,72–77], and of prevention, monitoring or mitigation of unwanted bias[23,28,33,40,47,52,54,58,64,69,73,74,78–80] and discrimination[28,33,36,38,44,45,50,55,56,60,68,81–84], the latter being significantly less referenced than the first two by the private sector. Whereas some sources focus on justice as respect for diversity[31,38,56,59,65,66,70,72,78,80,85,86], inclusion[31,45,47,51,72,80] and equality[41,45,51,59,60,72,78], others call for a possibility to appeal or challenge decisions[28,35–37,74,79], or the right to redress[33,42,45,46,50,68,85] and remedy[45,48]. Sources also emphasize the importance of fair access to AI[59,70,87], to data[33,37,44,67,83,88–90], and to the benefits of AI[37,38,80,91]. Issuers from the public sector place particular emphasis on AI's impact on the labor market[37,38,55,84,92], and the need to address democratic[33,38,59,73] or societal[31,48,55,65] issues. Sources focusing on the risk of biases within datasets underline the importance of acquiring and processing accurate, complete and diverse data[23,28,52,70,93], especially training data[27,33,35,38,52,58].

If specified, the preservation and promotion of justice are proposed to be pursued through: (a) technical solutions such as standards[50,68,89] or explicit normative encoding[28,37,43,67]; (b) transparency[54,62], notably by providing information[36,38,79] and raising public awareness of existing rights and regulation[28,59]; (c) testing[52,58,67,69], monitoring[54,56] and auditing[39,46,50,67], the preferred solution of notably data protection offices; (d) developing or strengthening the rule of law and the right to appeal, recourse, redress, or remedy[37,38,42,45,46,48,68,74,79]; (e) via systemic changes and processes such as governmental action[42,45,87,92] and oversight[94], a more interdisciplinary[47,65,85,93] or otherwise diverse[58,59,70,85,87,95] workforce, as well as better inclusion of civil society or other relevant stakeholders in an interactive manner[28,33,41,46,55,57,58,65,68,69,79,80,86] and increased attention to the distribution of benefits[25,33,38,48,63,76].

***Non-maleficence***

References to non-maleficence outweigh those to beneficence by a factor of 1.5 and encompass general calls for safety and security[80,90,96,97] or state that AI should never cause foreseeable or unintentional harm[23,30,33,56,60,79]. More granular considerations entail the avoidance of specific risks or potential harms, e.g. intentional misuse via cyberwarfare and malicious hacking[51,53,54,78,81,89], and suggest risk-management strategies. Harm is primarily interpreted as discrimination[38,44,47,48,50,95,98], violation of privacy[23,35,44,64,78,98,99], or bodily harm[25,30,31,33,56,92,96,100]. Less frequent characterizations include loss of trust[30] or skills[44], ‘radical individualism’[38], the risk that technological progress might outpace regulatory measures[57], negative impacts on long-term social well-being[44], on infrastructure[44], or on psychological[35,56], emotional[56] or economic aspects[44,56].

Harm-prevention guidelines focus primarily on technical measures and governance strategies, ranging from interventions at the level of AI research[27,47,64,79,85,101], design[23,25,27,32,39,56,58], technology development and/or deployment[54] to lateral and continuous approaches[33,55,63]. Technical solutions include in-built data quality evaluations[25] or security[23] and privacy by design[23,27,39], though notable exceptions also advocate for establishing industry standards[30,64,102]. Proposed governance strategies include active cooperation across disciplines and stakeholders[33,47,53,62], compliance with existing or new legislation[27,31,35,81,95,99], and the need to establish oversight processes and practices, notably tests[36,38,47,74,79], monitoring[36,58], audits and assessments by internal units, customers, users, independent third parties, or governmental entities[40,48,51,58,81,94,95,98], often geared towards standards for AI implementation and outcome assessment. Most sources explicitly mention potential 'dual-use'[8,32,33,38,60,79] or imply that damages may be unavoidable, in which case risks should be assessed[40,48,51], reduced[40,69,72–74], and mitigated[34,35,38,53,63,68], and the attribution of liability should be clearly defined[31,37,38,44,82].

***Responsibility and accountability***

Despite widespread references to 'responsible AI'[43,51,78,83], responsibility and accountability are rarely defined. Nonetheless, specific recommendations include acting with 'integrity'[47,52,60] and clarifying the attribution of responsibility and legal liability[23,58,78,103], if possible upfront[36], in contracts[52] or, alternatively, by centering on remedy[26]. In contrast, other sources suggest focusing on the underlying reasons and processes that may lead to potential harm[74,83]. Yet others underline the responsibility of whistleblowing in case of potential harm[36,55,60], and aim at promoting diversity[49,92] or introducing ethics into STEM education[59]. Very different actors are named as being responsible and accountable for AI's actions and decisions: AI developers[58,60,73,96], designers[36,44], 'institutions'[40,42] or 'industry'[69]. Further disagreement emerged on whether AI should be held accountable in a human-like manner[70] or whether humans should always be the only actors who are ultimately responsible for technological artifacts[31,32,35,37,52,92].

***Privacy***

Ethical AI sees privacy both as a value to uphold[44,64,75,99] and as a right to be protected[27,28,37,38,53]. While often undefined, privacy is often presented in relation to data protection[23,27,36,53,58,66,71,79,83,98] and data security[27,35,64,66,88,98]. A few sources link privacy to freedom[38,53] or trust[74,92]. Suggested modes of achievement fall into three categories: technical solutions[64,80] such as differential privacy[74,89], privacy by design[25,27,28,79,98], data minimization[36,58], and access control[36,58], calls for more research[47,64,74,98] and awareness[64,74], and regulatory approaches[25,52,71], with sources referring to legal compliance more broadly[27,32,36,58,60,81], or suggesting certificates[104] or the creation or adaptation of laws and regulations to accommodate the specificities of AI[64,74,88,105].

### *Beneficence*

While promoting good (*beneficence* in ethical terms) is often mentioned, it is rarely defined, though notable exceptions mention the augmentation of human senses[86], the promotion of human well-being and flourishing[34,90], peace and happiness[60], the creation of socio-economic opportunities[36], and economic prosperity[37,53]. Similar uncertainty concerns the actors that should benefit from AI: private sector issuers tend to highlight the benefit of AI for customers[23,48], though many sources require AI to be shared[49,52,76] and to benefit 'everyone'[36,59,65,84], 'humanity'[27,37,44,60,100,102], both of the above[48,66], 'society'[34,87], 'as many people as possible'[37,53,99], 'all sentient creatures'[83], the 'planet'[37,72] and the environment[38,90]. Strategies for the promotion of good include aligning AI with human values[34,44], advancing 'scientific understanding of the world'[100], minimizing power concentration[102] or, conversely, using power 'for the benefit of human rights'[82]; working more closely with 'affected' people[65], minimizing conflicts of interests[102]; proving beneficence through customer demand[48] and feedback[58], and developing new metrics and measurements for human well-being[44,90].

### *Freedom and autonomy*

Whereas some sources specifically refer to the freedom of expression[28,73,82,105] or informational self-determination[28,90] and 'privacy-protecting user controls'[58], others generally promote freedom[31,69,72], empowerment[28,52,99] or autonomy[31,33,62,77,81,96]. Some documents refer to autonomy as a positive freedom, specifically the freedom to flourish[36], to self-determination through democratic means[38], the right to establish and develop relationships with other human beings[38,92], the freedom to withdraw consent[67], or the freedom to use a preferred platform or technology[73,80]. Other documents focus on negative freedom, for example freedom from technological experimentation[82], manipulation[33] or surveillance[38]. Freedom and autonomy are believed to be promoted through transparency and predictable AI[38], by not 'reducing options for and knowledge of citizens'[38], by actively increasing people's knowledge about AI[36,52,62], giving notice and consent[79] or, conversely, by actively refraining from collecting and spreading data in absence of informed consent[30,38,44,55,74].

***Trust***

References to trust include calls for trustworthy AI research and technology[50,97,99], trustworthy AI developers and organisations[51,60,66], trustworthy 'design principles'[91], or underline the importance of customers' trust[23,52,58,66,74,80]. Calls for trust are proposed because a culture of trust among scientists and engineers is believed to support the achievement of other organisational goals[99], or because overall trust in the recommendations, judgments and uses of AI is indispensable for AI to 'fulfill its world changing potential'[24]. This last point is contradicted by one guideline explicitly warning against excessive trust in AI[81]. Suggestions for building or sustaining trust include education[33], reliability[50,51], accountability[56], processes to monitor and evaluate the integrity of AI systems over time[51] and tools and techniques ensuring compliance with norms and standards[43,63]. Whereas some guidelines require AI to be transparent[37,43,57,58], understandable[36,37], or explainable[52] in order to build trust, another one explicitly suggests that, instead of demanding understandability, it should be ensured that AI fulfills public expectations[50]. Other reported facilitators of trust include 'a Certificate of Fairness'[104], multi-stakeholder dialogue[64], awareness about the value of using personal data[74], and avoiding harm[30,56].

***Sustainability***

To the extent that is referenced, sustainability calls for development and deployment of AI to consider protecting the environment[33,38,46], improving the planet's ecosystem and biodiversity[37], contributing to fairer and more equal societies[65] and promoting peace[66]. Ideally, AI creates sustainable systems[44,76,90] that process data sustainably[43] and whose insights remain valid over time[48]. To achieve this aim, AI should be designed, deployed and managed with care[38] to increase its energy efficiency and minimize its ecological footprint[31]. To make future developments sustainable, corporations are asked to create policies ensuring accountability in the domain of potential job losses[37] and to use challenges as an opportunity for innovation[38].

***Dignity***

While dignity remains undefined in existing guidelines, safe the specification that it is a prerogative of humans but not robots[92], there is frequent reference to what it entails: dignity is intertwined with human rights[101] or otherwise means avoiding harm[31], forced acceptance[31], automated classification[38], and unknown human-AI interaction[38]. It is argued that AI should not diminish[33] or destroy[80] but respect[82], preserve[69] or even increase human dignity[36,37]. Dignity is believed to be preserved if it is respected by AI developers in the first place[96] and promoted through new legislation[38], through governance initiatives[36], or through government-issued technical and methodological guidelines[82].

### *Solidarity*

Solidarity is mostly referenced in relation to the implications of AI for the labor market[104]. Sources call for a strong social safety net[37,84]. They underline the need for redistributing the benefits of AI in order not to threaten social cohesion[49] and respecting potentially vulnerable persons and groups[33]. Lastly, there is a warning of data collection and practices focused on individuals which may undermine solidarity in favour of 'radical individualism'[38].

## Discussion

We found a rapid increase in the number and variety of guidance documents for ethical AI, demonstrating the increasing active involvement of the international community. Organisations publishing AI guidelines come from a wide range of sectors. In particular the nearly equivalent proportion of documents issued by the public sector (i.e. governmental and inter-governmental organisations) and the private sector (companies and private sector alliances) indicate that the ethical challenges of AI concern both public entities and private enterprises. However, there is significant divergence in the solutions proposed to meet the ethical challenges of AI. Further, the relative underrepresentation of geographic areas such as Africa, South and Central America and Central Asia indicates that the international debate over ethical AI may not be happening globally in equal measures. MEDC countries are shaping this debate more than others, which raises concerns about neglecting local knowledge, cultural pluralism and global fairness.

The proliferation of soft-law efforts can be interpreted as a governance response to advanced research into AI, whose research output and market size have drastically increased[106] in recent years. Our analysis shows the emergence of an apparent cross-stakeholder convergence on promoting the ethical principles of transparency, justice, non-maleficence, responsibility, and privacy. Nonetheless, our thematic analysis reveals substantive divergences in relation to four major factors: (i) how ethical principles are interpreted, (ii) why they are deemed important, (iii) what issue, domain or actors they pertain to, and (iv) how they should be implemented. Furthermore, unclarity remains as to which ethical principles should be prioritized, how conflicts between ethical principles should be resolved, who should enforce ethical oversight on AI and how researchers and institutions can comply with the resulting guidelines. These findings suggest the existence of a gap at the cross-section of principles formulation and their implementation into practice which can hardly be solved through technical expertise or top-down approaches.

Although no single ethical principle is explicitly endorsed by all existing guidelines, transparency, justice and fairness, non-maleficence, responsibility and privacy are each referenced in more than half of all guidelines. This focus could be indicating a developing convergence on ethical AI around these principles in the global policy landscape. In particular, the prevalence of calls for transparency, justice and fairness points to an emerging moral priority to require transparent processes throughout the entire AI continuum (from transparency in the development and design of algorithms to transparent practices for AI use), and to caution the global community against the risk that AI might increase inequality if justice and fairness considerations are not adequately addressed. Both these themes appear to be intertwined with the theme of responsibility, as the promotion of both transparency and justice seems to postulate increased responsibility and accountability on the side of AI makers and deployers.

It has been argued that transparency is not an ethical principle per se, but rather "a proethical condition for enabling or impairing other ethical practices or principles"[107]. The proethical nature of transparency might partly explain its higher prevalence compared to other ethical principles. It is notable that current guidelines place significant value in the promotion of responsibility and accountability, yet few of them emphasize the duty of all stakeholders involved in the development and deployment of AI to act with integrity. This mismatch is probably associated with the observation that existing guidelines fail to establish a full correspondence between principles and actionable requirements, with several principles remaining uncharacterized or disconnected from the requirements necessary for their realization.

As codes related to non-maleficence outnumber those related to beneficence, it appears that, for the current AI community, the moral obligation to preventing harm takes precedence over the promotion of good. This fact can be partly interpreted as an instance of the so-called negativity bias, i.e. a general cognitive bias to give greater weight to negative entities[108,109]. This negative characterization of ethical values is further emphasized by the fact that existing guidelines focus primarily on how to preserve privacy, dignity, autonomy and individual freedom *in spite of* advances in AI, while largely neglecting whether these principles could be promoted through responsible innovation in AI[110].

The issue of trust in AI, while being addressed by less than one third of all sources, tackles a critical ethical dilemma in AI governance: determining whether it is morally desirable to foster public trust in AI. While several sources, especially those produced within the private sector, highlight the importance of fostering trust in AI through educational and awareness-raising activities, a smaller number of sources contend that trust in AI may actually diminish scrutiny and undermine some societal obligations of AI producers[111]. This possibility would challenge the dominant view in AI ethics that building public trust in AI is a fundamental requirement for ethical governance[112].

The relative thematic underrepresentation of sustainability and solidarity suggests that these topics might be currently flying under the radar of the mainstream ethical discourse on AI. The underrepresentation of sustainability-related principles is particularly problematic in light of the fact that the deployment of AI requires massive computational resources which, in turn, require high energy consumption. The environmental impact of AI, however, does not only involve the negative effects of high-footprint digital infrastructures, but also the possibility of harnessing AI for the benefit of ecosystems and the entire biosphere. This latter point, highlighted in a report by the World Economic Forum[113] though not in the AI guidelines by the same institution, requires wider endorsement to become entrenched in the ethical AI narrative. The ethical principle of solidarity is sparsely referenced, typically in association with the development of inclusive strategies for the prevention of job losses and unfair sharing of burdens. Little attention is devoted to promoting solidarity through the emerging possibility of using AI expertise for solving humanitarian challenges, a mission that is currently being pursued, among others, by intergovernmental organisations such as the United Nations Office for Project Services (UNOPS)[114] or the World Health

Organization (WHO) and private companies such as Microsoft[115]. As the humanitarian cost of anthropogenic climate change is rapidly increasing[116], the principles of sustainability and solidarity appear strictly intertwined though poorly represented compared to other principles.

While numerical data indicate an emerging convergence around the promotion of some ethical principles, in-depth thematic analysis paints a more complicated picture, as there are critical differences in *how* these principles are interpreted as well as what requirements are considered to be necessary for their realization. Results show that different and often conflicting measures are proposed for the practical achievement of ethical AI. For example, the need for ever larger, more diverse datasets to "unbias" AI appears difficult to conciliate with the requirement to give individuals increased control over their data and its use in order to respect their privacy and autonomy. Similar contrasts emerge between the requirement of avoiding harm at all costs and that of balancing risks and benefits. Furthermore, it should be noted that risk-benefit evaluations will lead to different results depending on whose well-being it will be optimized for by which actors. If not resolved, such divergences and tensions may undermine attempts to develop a global agenda for ethical AI.

Despite a general agreement that AI should be ethical, significant divergences emerge within and between guidelines for ethical AI. Furthermore, uncertainty remains regarding how ethical principles and guidelines should be implemented. These challenges have implications for science policy, technology governance and research ethics. At the policy level, they urge increased cooperative efforts among governmental organisations to harmonize and prioritize their AI agendas, an effort that can be mediated and facilitated by inter-governmental organisations. While harmonization is desirable, however, it should not come at the costs of obliterating cultural and moral pluralism over AI. Therefore, a fundamental challenge for developing a global agenda for AI is balancing the need for cross-national harmonization over the respect for cultural diversity and moral pluralism. This challenge will require the development of deliberative mechanisms to adjudicate disagreement concerning the values and implications of AI advances among different stakeholders from different global regions. At the level of technology governance, harmonization is typically implemented in terms of standardizations. Efforts in this direction have been made, among others, by the Institute of Electrical and Electronics Engineers (IEEE) through the "Ethically Aligned Designed" initiative[117]. Finally, soft governance mechanisms such as Independent Review Boards (IRBs) will be increasingly required to assess the ethical validity of AI applications in scientific research, especially those in the academic domain. However, AI applications by governments or private corporations will unlikely fall under their oversight, unless significant expansions to the IRBs' purview are made.

The international community seems to converge on the importance of transparency, non-maleficence, responsibility, and privacy for the development and deployment of ethical AI. However, enriching the current ethical AI discourse through a better appraisal of critical yet underrepresented ethical principles such as human dignity, solidarity and sustainability is likely to result into a better articulated ethical landscape for artificial intelligence. Furthermore, shifting the focus from principle-formulation to translation into practice must be the next step. A global agenda for ethical AI should balance the need for cross-national and cross-domain harmonization over the respect for cultural diversity and moral pluralism. Overall, our review provides a useful starting point for understanding the inherent diversity of current principles and guidelines for ethical AI and outlines the challenges ahead for the global community.

## Limitations

This study has several limitations. First, guidelines and soft-law documents are an instance of gray literature, hence not indexed in conventional scholarly databases. Therefore, their retrieval is inevitably less replicable and unbiased compared to systematic database search of peer-reviewed literature. Following best practices for gray literature review, this limitation has been mitigated by developing a discovery and eligibility protocol which was pilot-tested prior to data collection. Although search results from search engines are personalized, the risk of personalization influencing discovery has been mitigated through the broadness of both the keyword search and the inclusion of results. A language bias may have skewed our corpus towards English results. Our content analysis presents the typical limitations of qualitative analytic methods. Following best practices for content analysis, this limitation has been mitigated by developing an inductive coding strategy which was conducted independently by two reviewers to minimize subjective bias. Finally, given the rapid pace of publication of AI guidance documents, there is a possibility that new policy documents were published after our search was completed. To minimize this risk, continuous monitoring of the literature was conducted in parallel with the data analysis and until April 23, 2019.

## Methods

We conducted a scoping review of the gray literature reporting principles and guidelines for ethical AI. A scoping review is a method aimed at synthesizing and mapping the existing literature[118], which is considered particularly suitable for complex or heterogeneous areas of research[118,119]. Given the absence of a unified database for AI-specific ethics guidelines, we developed a protocol for discovery and eligibility, adapted from the Preferred Reporting Items for Systematic Reviews and Meta-Analyses (PRISMA) framework[120]. The protocol was pilot-tested and calibrated prior to data collection. Following best practices for gray literature retrieval, a multi-stage screening strategy involving both inductive screening via search engine and deductive identification of relevant entities with associated websites and online collections was conducted. To achieve comprehensiveness and systematicity, relevant documents were retrieved by relying on three sequential search strategies (cf. Figure 2): First, a manual search of four link hub webpages ("linkhubs")[121–124] was performed. 68 sources were retrieved, out of which 30 were eligible (27 after removing duplicates). Second, a keyword-based web search of the Google.com search engine was performed in private browsing modus, after log-out from personal accounts and erasure of all web cookies and history.[125,126] Search was performed using the following keywords: [AI principles], [artificial intelligence principles], [AI guidelines], [artificial intelligence guidelines], [ethical AI] and [ethical artificial intelligence]. Every link in the first thirty search results was followed and screened (i) for AI principles, resulting in 10 more sources after removing duplicates, and (ii) for articles mentioning AI principles, leading to the identification of 3 additional non-duplicate sources. The remaining Google results up to the 200th listings for each Google search were followed and screened for AI principles only. Within these additional 1020 link listings we identified 15 non-duplicate documents. After identifying relevant documents through the two processes above, we used citation-chaining to manually screen the full-texts and, if applicable, reference lists of all eligible sources in order to identify other relevant documents. 17 additional sources were identified. We continued to monitor the literature in parallel with the data analysis and until April 23, 2019, to retrieve eligible documents that were released after our search was completed. Twelve new sources were included within this extended time frame. To ensure theoretical saturation, we exhausted the citation chaining within all identified sources until no additional relevant document could be identified.

*Figure 2- PRISMA-based flowchart of retrieval process*

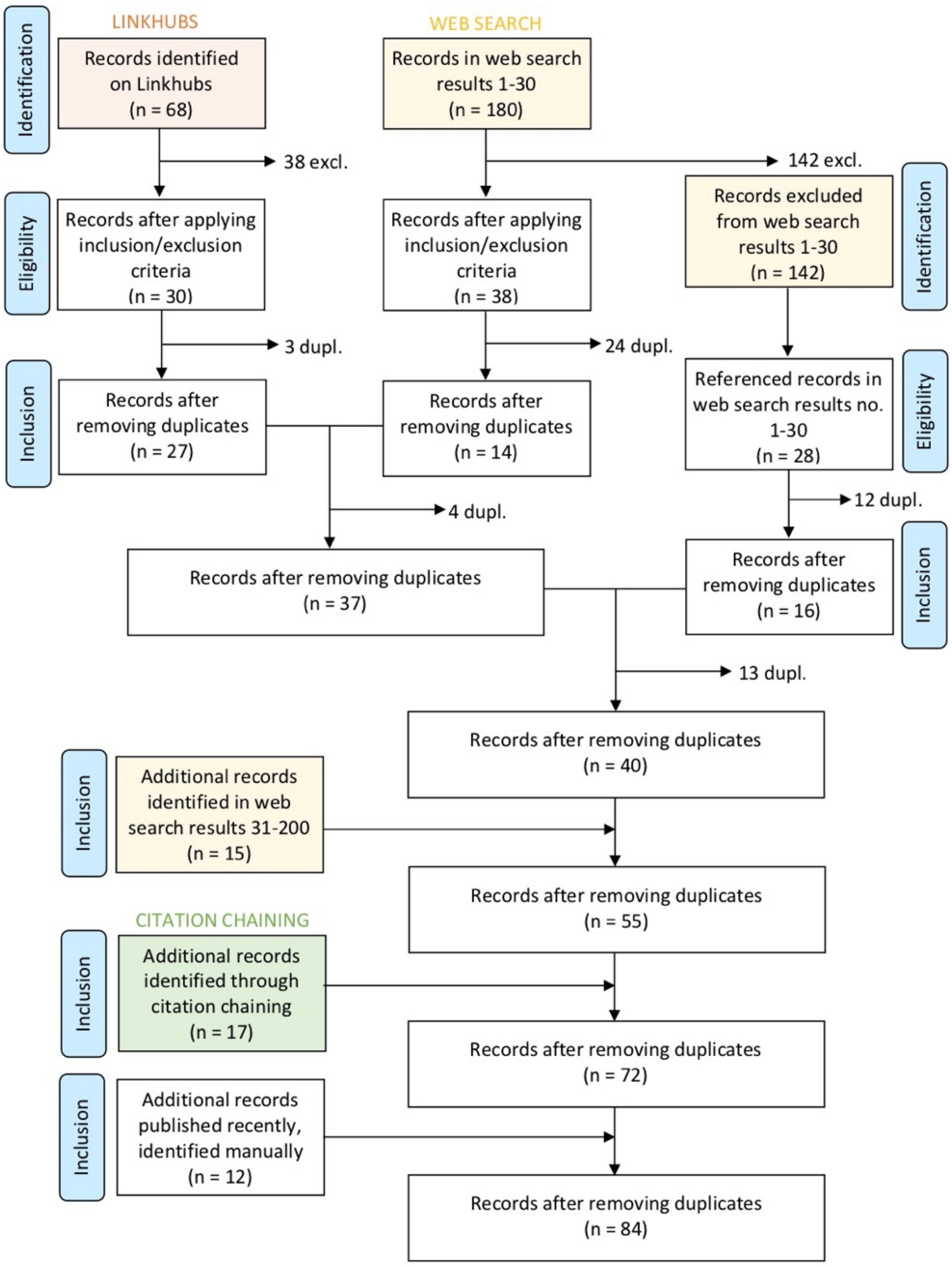

Flowchart of our retrieval process based on the PRISMA template for systematic reviews[127]. We relied on three search strategies (linkhubs, web search and citation chaining) and added the most recent records manually, identifying a total of 84 eligible, non-duplicate documents containing ethical principles for AI.

Based on our inclusion/exclusion criteria, policy documents (including principles, guidelines and institutional reports) included in the final synthesis were (i) written in English, German, French, Italian, Greek; (ii) issued by institutional entities from both the public and the public sectors; (iii) referred explicitly in their title/description to AI or ancillary notions, (iv) expressed a normative ethical stance defined as a moral preference for a defined course of action (cf. SI Table S2). Following full-text screening, 84 sources or parts thereof were included in the final synthesis (cf. SI Table S1).

Content analysis of the 84 sources was independently conducted by two researchers in two cycles of manual coding and one cycle of code mapping within the qualitative data analysis software *Nvivo* for Mac v.11.4. During the first cycle of coding, one researcher exhaustively tagged all relevant text through inductive coding[128] attributing a total of 3457 codes, out of which 1180 were subsequently discovered to pertain to ethical principles. Subsequently, two researchers conducted the code mapping process in order to reduce subjective bias. The process of code mapping, a method for qualitative metasynthesis[129], consisted of two iterations of themeing[128], whereby categories were first attributed to each code, then categorized in turn (cf. SI Table S3). For the theming of ethical principles, we relied deductively on normative ethical literature. Ethical categories were inspected and assessed for consistency by two researchers with primary expertise in ethics. Thirteen ethical categories emerging from code mapping, two of which were merged with others due to independently assessed semantic and thematic proximity. Finally, we extracted significance and frequency by applying focused coding, a second cycle coding methodology used for interpretive analysis[128], to the data categorized in ethical categories. Consistency check was performed both by reference to the relevant ethical literature and a process of deliberative mutual adjustment among the general principles and the particular judgments contained in the policy documents, an analytic strategy known as 'reflective equilibrium'[130].

## REFERENCES

1. Harari, Y. N. Reboot for the AI revolution. *Nat. News* **550**, 324 (2017).
2. Appenzeller, T. The AI revolution in science. *Science* (2017). doi:10.1126/science.aan7064
3. Jordan, M. I. & Mitchell, T. M. Machine learning: Trends, perspectives, and prospects. *Science* **349**, 255–260 (2015).
4. Stead, W. W. Clinical Implications and Challenges of Artificial Intelligence and Deep Learning. *JAMA* **320**, 1107–1108 (2018).
5. Vayena, E., Blasimme, A. & Cohen, I. G. Machine learning in medicine: Addressing ethical challenges. *PLOS Med.* **15**, e1002689 (2018).
6. Awad, E. *et al.* The Moral Machine experiment. *Nature* **563**, 59 (2018).
7. Science must examine the future of work. *Nat. News* **550**, 301 (2017).
8. Brundage, M. *et al. The Malicious Use of Artificial Intelligence: Forecasting, Prevention, and Mitigation.* (Future of Humanity Institute; University of Oxford; Centre for the Study of Existential Risk; University of Cambridge; Center for a New American Security; Electronic Frontier Foundation; OpenAI, 2018).
9. Zou, J. & Schiebinger, L. AI can be sexist and racist — it's time to make it fair. *Nature* **559**, 324 (2018).
10. Boddington, P. *Towards a Code of Ethics for Artificial Intelligence*. (Springer International Publishing, 2017).
11. Bostrom, N. & Yudkowsky, E. The ethics of artificial intelligence. in *The Cambridge Handbook of Artificial Intelligence* (eds. Frankish, K. & Ramsey, W. M.) 316–334 (Cambridge University Press, 2014). doi:10.1017/CBO9781139046855.020
12. Etzioni, A. & Etzioni, O. AI Assisted Ethics. *Ethics Inf. Technol.* **18**, 149–156 (2016).

13. Yuste, R. *et al.* Four ethical priorities for neurotechnologies and AI. *Nat. News* **551**, 159 (2017).

14. Cath, C., Wachter, S., Mittelstadt, B., Taddeo, M. & Floridi, L. Artificial Intelligence and the 'Good Society': the US, EU, and UK approach. *Sci. Eng. Ethics* **24**, 505–528 (2018).

15. Zeng, Y., Lu, E. & Huangfu, C. Linking Artificial Intelligence Principles. *ArXiv181204814 Cs* (2018).

16. Greene, D., Hoffmann, A. L. & Stark, L. Better, Nicer, Clearer, Fairer: A Critical Assessment of the Movement for Ethical Artificial Intelligence and Machine Learning. in (2019).

17. Crawford, K. & Calo, R. There is a blind spot in AI research. *Nat. News* **538**, 311 (2016).

18. Altman, M., Wood, A. & Vayena, E. A Harm-Reduction Framework for Algorithmic Fairness. *IEEE Secur. Priv.* **16**, 34–45 (2018).

19. Bolukbasi, T., Chang, K.-W., Zou, J., Saligrama, V. & Kalai, A. Man is to Computer Programmer as Woman is to Homemaker? Debiasing Word Embeddings. *ArXiv160706520 Cs Stat* (2016).

20. O'Neil, C. *Weapons of Math Destruction: How Big Data Increases Inequality and Threatens Democracy*. (Crown, 2016).

21. Veale, M. & Binns, R. Fairer machine learning in the real world: Mitigating discrimination without collecting sensitive data. *Big Data Soc.* **4**, 205395171774353 (2017).

22. Wagner, B. Ethics as an escape from regulation. From "ethics-washing" to ethics-shopping? in *Being profiled: cogitas ergo sum : 10 years of 'profiling the European* *citizen'* (eds. Bayamlioglu, E., Baraliuc, I., Janssens, L. A. W. & Hildebrandt, M.) 84–89 (Amsterdam University Press, 2018).

23. Deutsche Telekom. Deutsche Telekom's guidelines for artificial intelligence. (2018).
24. IBM. Transparency and Trust in the Cognitive Era. *IBM* (2017). Available at: https://www.ibm.com/blogs/think/2017/01/ibm-cognitive-principles/. (Accessed: 21st February 2019)
25. Initial code of conduct for data-driven health and care technology. *GOV.UK* Available at: https://www.gov.uk/government/publications/code-of-conduct-for-data-driven-health-and-care-technology/initial-code-of-conduct-for-data-driven-health-and-care-technology. (Accessed: 1st November 2018)
26. Diakopoulos, N. *et al.* Principles for Accountable Algorithms. *FATML | Principles for Accountable Algorithms and a Social Impact Statement for Algorithms* (2016). Available at: http://www.fatml.org/resources/principles-for-accountable-algorithms. (Accessed: 21st February 2019)
27. Telefónica. Our Artificial Intelligence Principles. (2018).
28. Commission Nationale de l'Informatique et des Libertés (CNIL), European Data Protection Supervisor (EDPS) & Garante per la protezione dei dati personali. Declaration on Ethics and Data Protection in Artificial Intelligence. (2018).
29. IBM. Everyday Ethics for Artificial Intelligence. A practical guide for designers & developers. (2018).
30. Federal Ministry of Transport and Digital Infrastructure, Ethics Commission. BMVI - Ethics Commission's complete report on automated and connected driving. (2017).
31. Green Digital Working Group. Position on Robotics and Artificial Intelligence. (2016).

32. EPSRC. Principles of robotics. *Engineering and Physical Sciences Research Council UK (EPSRC)* (2011). Available at: https://epsrc.ukri.org/research/ourportfolio/themes/engineering/activities/principlesofrobotics/. (Accessed: 21st February 2019)
33. High-Level Expert Group on Artificial Intelligence. Ethics Guidelines for Trustworthy AI. (2019).
34. Dubai. AI Principles & Ethics. *Smart Dubai* (2019). Available at: http://www.smartdubai.ae/initiatives/ai-principles-ethics. (Accessed: 8th April 2019)
35. Dawson, D. *et al.* Artificial Intelligence: Australia's Ethics Framework. (2019).
36. Internet Society. Artificial Intelligence & Machine Learning: Policy Paper. *Internet Society* (2017). Available at: https://www.internetsociety.org/resources/doc/2017/artificial-intelligence-and-machine-learning-policy-paper/. (Accessed: 21st February 2019)
37. UNI Global. 10 Principles for Ethical AI. (2017).
38. European Group on Ethics in Science and New Technologies. Statement on Artificial Intelligence, Robotics and 'Autonomous' Systems. (2018).
39. Information Commissioner's Office UK. Big data, artificial intelligence, machine learning and data protection. (2017).
40. The Public Voice. Universal Guidelines for Artificial Intelligence. *The Public Voice* (2018). Available at: https://thepublicvoice.org/ai-universal-guidelines/. (Accessed: 21st February 2019)
41. The Future Society. Science, Law and Society (SLS) Initiative. *The Future Society* (2018). Available at: https://web.archive.org/web/20180621203843/http://thefuturesociety.org/science-law-society-sls-initiative/. (Accessed: 25th February 2019)

42. Association for Computing Machinery (ACM). Statement on Algorithmic Transparency and Accountability. (2017).

43. Special Interest Group on Artificial Intelligence. Dutch Artificial Intelligence Manifesto. (2018).

44. Ethically Aligned Design. A Vision for Prioritizing Human Well-being with Autonomous and Intelligent Systems. Ethically Aligned Design. A Vision for Prioritizing Human Well-being with Autonomous and Intelligent Systems V.2. (2017).

45. Access Now. The Toronto Declaration: Protecting the right to equality and non-discrimination in machine learning systems. (2018).

46. Floridi, L. *et al. AI4People—An Ethical Framework for a Good AI Society: Opportunities, Risks, Principles, and Recommendations*. (AI4People).

47. SAP. SAP's guiding principles for artificial intelligence (AI). *SAP* (2018). Available at: https://www.sap.com/products/leonardo/machine-learning/ai-ethics.html#guiding-principles. (Accessed: 19th February 2019)

48. Software & Information Industry Association (SIIA), Public Policy Division. Ethical Principles for Artificial Intelligence and Data Analytics. (2017).

49. Koski, O. & Husso, K. Work in the age of artificial intelligence. (2018).

50. Center for Democracy & Technology. Digital Decisions. *Center for Democracy & Technology* Available at: https://cdt.org/issue/privacy-data/digital-decisions/. (Accessed: 21st February 2019)

51. MI Garage. Ethics Framework - Responsible AI. *MI Garage* Available at: https://www.migarage.ai/ethics-framework/. (Accessed: 22nd February 2019)

52. Institute of Business Ethics. Business Ethics and Artificial Intelligence. (2018).

53. Asilomar AI Principles. *Future of Life Institute* (2017). Available at: https://futureoflife.org/ai-principles/. (Accessed: 1st November 2018)

54. PricewaterhouseCoopers. The responsible AI framework. *PwC* Available at: https://www.pwc.co.uk/services/audit-assurance/risk-assurance/services/technology-risk/technology-risk-insights/accelerating-innovation-through-responsible-ai/responsible-ai-framework.html. (Accessed: 22nd February 2019)

55. Whittaker, M. *et al.* AI Now Report 2018. (2018).

56. Personal Data Protection Commission Singapore. Discussion Paper on AI and Personal Data -- Fostering Responsible Development and Adoption of AI. (2018).

57. Royal College of Physicians. Artificial Intelligence (AI) in Health. *RCP London* (2018). Available at: https://www.rcplondon.ac.uk/projects/outputs/artificial-intelligence-ai-health.

58. Microsoft. Responsible bots: 10 guidelines for developers of conversational AI. (2018).

59. Villani, C. *For a meaningful Artificial Intelligence. Towards a French and European strategy*. (Mission assigned by the Prime Minister Édouard Philippe, 2018).

60. The Japanese Society for Artificial Intelligence. The Japanese Society for Artificial Intelligence Ethical Guidelines. (2017).

61. Demiaux, V. How can humans keep the upper hand? The ethical matters raised by algorithms and artificial intelligence. (2017).

62. Council of Europe: CEPEJ. European ethical Charter on the use of Artificial Intelligence in judicial systems and their environment. (2019).

63. American College of Radiology *et al.* Ethics of AI in Radiology: European and North American Multisociety Statement. (2019).

64. Leaders of the G7. Charlevoix Common Vision for the Future of Artificial Intelligence. (2018).

65. DeepMind Ethics&Society. DeepMind Ethics & Society Principles. *DeepMind* (2017). Available at: https://deepmind.com/applied/deepmind-ethics-society/principles/. (Accessed: 21st February 2019)

66. Sony. Sony Group AI Ethics Guidelines. (2018).

67. Datatilsynet. *Artificial intelligence and privacy*. (The Norwegian Data Protection Authority, 2018).

68. WEF. White Paper: How to Prevent Discriminatory Outcomes in Machine Learnig. (2018).

69. Information Technology Industry Council (ITI). ITI AI Policy Principles. (2017).

70. Sage. The Ethics of Code: Developing AI for Business with Five Core Principles. (2017).

71. OP Group. Commitments and principles. *OP* Available at: https://www.op.fi/op-financial-group/corporate-social-responsibility/commitments-and-principles. (Accessed: 21st February 2019)

72. Tieto. Tieto's AI ethics guidelines. (2018).

73. Unity. Introducing Unity's Guiding Principles for Ethical AI – Unity Blog. *Unity Technologies Blog* (2018).

74. National Institution for Transforming India (Niti Aayog). Discussion Paper: National Strategy for Artificial Intelligence. (2018).

75. House of Lords. *AI in the UK: ready, willing and able*. 183 (2018).

76. The Information Accountability Foundation. Unified Ethical Frame for Big Data Analysis IAF Big Data Ethics Initiative, Part A. (2015).

77. Fenech, M., Nika Strukelj & Olly Buston. *Ethical, social, and political challenges of Artificial Intelligence in Health*. (Future Advocacy, 2019).

78. Accenture UK. Responsible AI and robotics. An ethical framework. Available at: https://www.accenture.com/gb-en/company-responsible-ai-robotics. (Accessed: 22nd February 2019)

79. Google. Our Principles. *Google AI* (2018). Available at: https://ai.google/principles/. (Accessed: 19th February 2019)

80. Microsoft. Microsoft AI principles. *Our approach* (2017). Available at: https://www.microsoft.com/en-us/ai/our-approach-to-ai. (Accessed: 1st November 2018)

81. CERNA Commission de réflexion sur l'Éthique de la Rechercheen sciences et technologies du Numérique d'Allistene. *Éthique de la rechercheen robotique*. (Allistene, 2014).

82. Est, R. van & Gerritsen, J. *Human rights in the robot age: Challenges arising from the use of robotics, artificial intelligence, and virtual and augmented reality*. (The Rathenau Institute, 2017).

83. Université de Montréal. Montreal Declaration. *The Declaration - Montreal Responsible AI* (2017). Available at: https://www.montrealdeclaration-responsibleai.com/the-declaration. (Accessed: 21st February 2019)

84. Government of the Republic of Korea. Mid- to Long-Term Master Plan in Preparation for the Intelligent Information Society. Managing the Fourth Industrial Revolution. (2017).

85. Crawford, K. *et al.* The AI Now Report. The Social and Economic Implications of Artificial Intelligence Technologies in the Near-Term. (2016).

86. Advisory Board on Artificial Intelligence and Human Society. *Report on Artificial Intelligence and Human Society Unofficial translation*. (Ministry of State for Science and Technology Policy, 2017).

87. Executive Office of the President; National Science and Technology Council; Committee on Technology. Preparing for the future of Artificial Intelligence. (2016).

88. Intel. Artificial intelligence. The public policy opportunity. (2017).

89. Royal Society. *Machine learning: the power and promise of computers that learn by example.* (Royal Society (Great Britain), 2017).

90. IEEE Global Initiative on Ethics of Autonomous and Intelligent Systems. Ethically Aligned Design: A Vision for Prioritizing Human Well-being with Autonomous and Intelligent Systems, First Edition (EAD1e). (2019).

91. European Parliament. *Report with recommendations to the Commission on Civil Law Rules on Robotics*. (2017).

92. COMEST/UNESCO. Report of COMEST on robotics ethics. *UNESDOC Digital Library* (2017). Available at: https://unesdoc.unesco.org/ark:/48223/pf0000253952. (Accessed: 21st February 2019)

93. Campolo, A., Madelyn Sanfilippo, Meredith Whittaker & Kate Crawford. AI Now 2017 Report. (2017).

94. American Medical Association (AMA). Policy Recommendations on Augmented Intelligence in Health Care H-480.940. (2018). Available at: https://policysearch.ama-assn.org/policyfinder/detail/AI?uri=%2FAMADoc%2FHOD.xml-H-480.940.xml. (Accessed: 21st February 2019)

95. Avila, R., Ana Brandusescu, Juan Ortiz Freuler & Dhanaraj Thakur. Artificial Intelligence: open questions about gender inclusion. (2018).

96. The Conference toward AI Network Society. Draft AI R&D Guidelines for International Discussions. (2017). Available at: http://www.soumu.go.jp/main_content/000507517.pdf. (Accessed: 21st February 2019)
97. National Science and Technology Council; Networking and Information Technology Research and Development Subcommittee. The National Artificial Intelligence Research and Development Strategic Plan. (2016).
98. Hoffmann, D. & Masucci, R. Intel's AI Privacy Policy White Paper. Protecting individuals' privacy and data in the artificial intelligence world. (2018).
99. Partnership on AI. Tenets. *The Partnership on AI* (2016). Available at: https://www.partnershiponai.org/tenets/. (Accessed: 21st February 2019)
100. Icelandic Institute for Intelligent Machines (IIIM). Ethics Policy. *IIIM* (2015). Available at: http://www.iiim.is/2015/08/ethics-policy/. (Accessed: 21st February 2019)
101. Latonero, M. *Governing Artificial Intelligence. Upholding Human Rights & Dignity*. (Data & Society, 2018).
102. OpenAI. OpenAI Charter. *OpenAI* (2018). Available at: https://blog.openai.com/openai-charter/. (Accessed: 21st February 2019)
103. Agenzia per l'Italia Digitale (AGID). L'intelligenzia artificiale al servizio del cittadino. (2018).
104. Women Leading in AI. *10 Principles of responsible AI.* (2019).
105. Privacy International & Article 19. Privacy and Freedom of Expression In the Age of Artificial Intelligence. (2018).
106. Shoham, Y. *et al. The AI Index 2018 Annual Report*. (AI Index Steering Committee, Human-Centered AI Initiative, Stanford University, 2018).

107. Turilli, M. & Floridi, L. The ethics of information transparency. *Ethics Inf. Technol.* **11**, 105–112 (2009).

108. Rozin, P. & Royzman, E. B. Negativity Bias, Negativity Dominance, and Contagion: *Personal. Soc. Psychol. Rev.* (2016). doi:10.1207/S15327957PSPR0504_2

109. Bentley, P. J., Brundage, M., Häggström, O. & Metzinger, T. *Should we fear artificial intelligence?: in-depth analysis*. (European Parliamentary Research Service: Scientific Foresight Unit (STOA), 2018).

110. Taddeo, M. & Floridi, L. How AI can be a force for good. *Science* **361**, 751–752 (2018).

111. Bryson, J. AI & Global Governance: No One Should Trust AI - Centre for Policy Research at United Nations University. *United Nations University. Centre for Policy Research* (2018). Available at: https://cpr.unu.edu/ai-global-governance-no-one-should-trust-ai.html. (Accessed: 21st March 2019)

112. Winfield, A. F. T. & Marina, J. Ethical governance is essential to building trust in robotics and artificial intelligence systems. *Philos. Trans. R. Soc. Math. Phys. Eng. Sci.* **376**, 20180085 (2018).

113. WEF. *Harnessing Artificial Intelligence for the Earth*. (WEF, 2018).

114. Lancaster, C. Can artificial intelligence improve humanitarian responses? *UNOPS* (2018). Available at: https://www.unops.org/news-and-stories/insights/can-artificial-intelligence-improve-humanitarian-responses. (Accessed: 22nd March 2019)

115. Microsoft. AI for Humanitarian Action. *Microsoft | AI* Available at: https://www.microsoft.com/en-us/ai/ai-for-humanitarian-action. (Accessed: 22nd March 2019)

116. Scheffran, J., Brzoska, M., Kominek, J., Link, P. M. & Schilling, J. Climate change and violent conflict. *Science* **336**, 869–871 (2012).

117. IEEE. The IEEE Global Initiative on Ethics of Autonomous and Intelligent Systems. *IEEE Standards Association* Available at: https://standards.ieee.org/industry-connections/ec/autonomous-systems.html. (Accessed: 22nd March 2019)

118. Arksey, H. & O'Malley, L. Scoping studies: towards a methodological framework. *Int. J. Soc. Res. Methodol.* **8**, 19–32 (2005).

119. Pham, M. T. *et al.* A scoping review of scoping reviews: advancing the approach and enhancing the consistency. *Res. Synth. Methods* **5**, 371–385 (2014).

120. Liberati, A. *et al.* The PRISMA Statement for Reporting Systematic Reviews and Meta-Analyses of Studies That Evaluate Health Care Interventions: Explanation and Elaboration. *PLOS Med.* **6**, e1000100 (2009).

121. Boddington, P. Alphabetical List of Resources. *Ethics for Artificial Intelligence* (2018). Available at: https://www.cs.ox.ac.uk/efai/resources/alphabetical-list-of-resources/. (Accessed: 4th May 2019)

122. Winfield, A. Alan Winfield's Web Log: A Round Up of Robotics and AI ethics. *Alan Winfield's Web Log* (2017).

123. Future of Life Institute. National and International AI Strategies. *Future of Life Institute* (2018). Available at: https://futureoflife.org/national-international-ai-strategies/. (Accessed: 4th May 2019)

124. Future of Life Institute. Summaries of AI Policy Resources. *Future of Life Institute* (2018). Available at: https://futureoflife.org/ai-policy-resources/. (Accessed: 4th May 2019)

125. Hagstrom, C., Kendall, S. & Cunningham, H. Googling for grey: using Google and Duckduckgo to find grey literature. in *Abstracts of the 23rd Cochrane Colloquium* **10 (Suppl): LRO 3.6**, 40 (Cochrane Database of Systematic Reviews, 2015).

126. Piasecki, J., Waligora, M. & Dranseika, V. Google Search as an Additional Source in Systematic Reviews. *Sci. Eng. Ethics* (2017). doi:10.1007/s11948-017-0010-4

127. Moher, D., Liberati, A., Tetzlaff, J., Altman, D. G. & Group, T. P. Preferred Reporting Items for Systematic Reviews and Meta-Analyses: The PRISMA Statement. *PLOS Med.* **6**, e1000097 (2009).

128. Saldaña, J. *The coding manual for qualitative researchers*. (Sage, 2013).

129. Noblit, G. W. & Hare, R. D. *Meta-Ethnography: Synthesizing Qualitative Studies*. (SAGE, 1988).

130. Daniels, N. *Justice and Justification: Reflective Equilibrium in Theory and Practice*. (Cambridge University Press, 1996).

# Supplementary Information for

## Artificial Intelligence: the global landscape of ethics guidelines

Anna Jobin [a], Marcello Ienca [a], Effy Vayena [a]*

[a] Health Ethics & Policy Lab, ETH Zurich, 8092 Zurich, Switzerland

* Corresponding author: effy.vayena@hest.ethz.ch

**This PDF file includes:**

Tables S1 to S3

## Table S1. Ethics guidelines for AI by date of publishing (incl. details)

| Name of Document/Website | Name of guidelines/principles | Issuer | Country of issuer | Type of issuer | Date of publishing | Target audience | Retrieval |
|---|---|---|---|---|---|---|---|
| Principles of robotics | Principles for designers, builders and users of robots | Engineering and Physical Sciences Research Council UK (EPSRC) | UK | Science foundation | 1-Apr-2011 | multiple (public, developers) | Linkhubs |
| Ethique de la recherche en robotique | Préconisations | CERNA (Allistene) | France | Research alliance | xx-Nov-2014 | researchers | Citation chaining |
| Unified Ethical Frame for Big Data Analysis. IAF Big Data Ethics Initiative, Part A | Values for an Ethical Frame | The Information Accountability Foundation | UK | NPO/Charity | xx-Mar-2015 | unspecified | Citation chaining |
| Ethics Policy | IIIM's Ethics Policy | Icelandic Institute for Intelligent Machines (IIIM) | Iceland | Academic and research institution | 31-Aug-2015 | self | Linkhubs |
| The AI Now Report. The Social and Economic Implications of Artificial Intelligence Technologies in the Near-Term | Key recommendations | AI Now Institute | USA | Academic and research institution | 22-Sep-2016 | unspecified | Citation chaining |
| Tenets | Tenets | Partnership on AI | n.a. | Private sector alliance | 29-Sep-2016 | self | Web search results 1-30 |
| Preparing for the future of Artificial Intelligence | Recommendations in this Report | Executive Office of the President; National Science and Technology Council; Committee on Technology | USA | Governmental agencies/organizations | xx-Oct-2016 | multiple (stakeholders engaged at variouspoints in the production, use, governance, and assessment of AI systems) | Linkhubs |
| The National Artificial Intelligence Research and Development Strategic Plan | R&D Strategy | National Science and Technology Council; Networking and Information Technology Research and Development Subcommittee | USA | Governmental agencies/organizations | xx-Oct-2016 | self | Linkhubs |
| Position on Robotics and Artificial Intelligence | 3. Principles // 6. Recommendations Green position on Robotics and Artificial Intelligence | The Greens (Green Working Group Robots) | EU | Political Party | 22-Nov-2016 | multiple (EU parliament, public, self) | Web search results 31-200 |
| Principles for Accountable Algorithms and a Social Impact Statement for Algorithms | Principles for Accountable Algorithms | Fairness, Accountability, and Transparency in Machine Learning (FATML) | n.a. | n.a. | 24-Nov-2016 | multiple (developers and product managers) | Linkhubs |
| Statement on Algorithmic Transparency and Accountability | Principles for Algorithmic Transparency and Accountability | Association for Computing Machinery (ACM) | USA | Prof. Association/Society | 12-Jan-2017 | multiple (developers, deployers) | Linkhubs |
| Report with recommendations to the Commission on Civil Law Rules on Robotics | Motion for a European Parliament Resolution | European Parliament | EU | IGO/supra-national | 27-Jan-2017 | public sector (lawmakers) | Linkhubs |
| AI Principles | AI Principles | Future of Life Institute | USA | Miscellaneous (mixed crowdsourced, NPO) | 30-Jan-2017 | unspecified | Linkhubs |
| The Japanese Society for Artificial Intelligence Ethical Guidelines | The Japanese Society for Artificial Intelligence Ethical Guidelines | Japanese Society for Artificial Intelligence | Japan | Prof. Association/Society | 28-Feb-2017 | self (incl AI) | Linkhubs |
| Report on Artificial Intelligence and Human Society (Unofficial translation) | 4.1 Ethical issues | Advisory Board on Artificial Intelligence and Human Society (initiative of the Minister of State for Science and Technology Policy) | Japan | Governmental agencies/organizations | 24-Mar-2017 | multiple (researchers, government, businesses, public, educators) | Web search results 31-200 |
| Artificial Intelligence and Machine Learning: Policy Paper | Guiding principles and recommendations | Internet Society | international | NPO/charity | 18-Apr-2017 | multiple (policymakers, other stakeholders in the wider Internet ecosystem) | Web search results 31-200 |
| Machine learning: the power and promise of computers that learn by example | Chapter six – A new wave of machine learning research | The Royal Society | UK | Prof. Association/Society | xx-Apr-2017 | unspecified | Citation chaining |
| The Ethics of Code: Developing AI for Business with Five Core Principles | The Ethics of Code: Developing AI for Business with Five Core Principles | Sage | UK | Company | 27-Jun-2017 | self | Citation chaining |
| Automated and Connected Driving: Report | Ethical rules for automated and connected vehicular traffic | Federal Ministry of Transport and Digital Infrastructure, Ethics Commission | Germany | Governmental agencies/organizations | xx-Jun-2017 | multiple (automated & connected vehicular traffic) | Linkhubs |
| Mid- to Long-Term Master Plan in Preparation for the Intelligent Information Society | Tasks (8-12) | Government of the Republic of Korea | South Korea | Governmental agencies/organizations | 20-Jul-2017 | self (gov) | Linkhubs |
| Draft AI R&D Guidelines for International Discussions | AI R&D Principles | Institute for Information and Communications Policy (IICP), The Conference toward AI Network Society | Japan | Governmental agencies/organizations | 28-Jul-2017 | multiple (systems and developers) | Linkhubs |
| Big data, artificial intelligence, machine learning and data protection | Key recommendations | Information Commissioner's Office | UK | Gov | 4-Sep-2017 | organisations | Web search results 1-30 |
| Report of COMEST on Robotics Ethics (only section "Recommendations" taken into account) | Relevant ethical principles and values | COMEST/UNESCO | international | IGO/supra-national | 14-Sep-2017 | unspecified | Citation chaining |
| Ethical Principles for Artificial Intelligence and Data Analytics | Ethical Principles for Artificial Intelligence and Data Analytics | Software & Information Industry Association (SIIA), Public Policy Division | international | Private sector alliance | 15-Sep-2017 | private sector (industry organizations) | |
| AI - Our approach | AI - Our approach | Microsoft | USA | Company | 7-Oct-2017 | self | Web search results 1-30 |
| DeepMind Ethics & Society Principles | Our Five Core Principles | DeepMind Ethics & Society | UK | Company | 10-Oct-2017 | self | Citation chaining |
| Human Rights in the Robot Age Report | Recommendations | The Rathenau Institute | Netherlands | Academic and research institution (Gov) | 11-Oct-2017 | public sector (Council of Europe) | Citation chaining |
| Artificial Intelligence. The Public Policy Opportunity | Summary of Recommendations | Intel Corporation | USA | Company | 18-Oct-2017 | public sector (policy makers) | Citation chaining |
| ITI AI Policy Principles | ITI AI Policy Principles | Information Technology Industry Council (ITI) | international | Private sector alliance | 24-Oct-2017 | self (members) | Citation chaining |
| AI Now 2017 Report | Recommendations, Executive Summary | AI Now Institute | USA | Academic and research institution | xx-Oct-2017 | multiple (core public agencies, companies, industry, universities, conferences, other stakeholders) | Citation chaining |
| Montréal Declaration: Responsible AI | Montréal Declaration: Responsible AI | Université de Montréal | Canada | Academic and research institution | 3-Nov-2017 | multiple (public, developers, policy makers) | Linkhubs |
| Ethically Aligned Design. A Vision for Prioritizing Human Well-being with Autonomous and Intelligent Systems, version 2 | Ethically Aligned Design. A Vision for Prioritizing Human Well-being with Autonomous and Intelligent Systems, version 2 | Institute of Electrical and Electronics Engineers (IEEE), The IEEE Global Initiative on Ethics of Autonomous and Intelligent Systems | international | Prof. Association/Society | 12-Dec-2017 | unspecified | Linkhubs |
| How can humans keep the upper hand? Report on the ethical matters raised by AI algorithms (only section "From principles to policy recommendations") | From principles to policy recommendations | French Data Protection Authority (CNIL) | France | Governmental agencies/organizations | 15-Dec-2017 | unspecified | Linkhubs |
| Top 10 Principles for Ethical Artificial Intelligence | Top 10 Principles for Ethical Artificial Intelligence | UNI Global Union | international | Federation/Union | 17-Dec-2017 | multiple (unions, workers) | Linkhubs |
| Business Ethics and Artificial Intelligence | Fundamental Values and Principles | Institute of Business Ethics | UK | Private sector alliance | 11-Jan-2018 | private sector (users of AI in business) | Web search results 31-200 |
| IBM's Principles for Trust and Transparency | IBM's Principles for Trust and Transparency | IBM | USA | Company | 17-Jan-2018 | self | Web search results 1-30 |
| Artificial intelligence and privacy | Recommendations for privacy friendly development and use of AI | The Norwegian Data Protection Authority | Norway | Governmental agencies/organizations | xx-Jan-2018 | multiple (developers, system suppliers, organisations, end users, authorities) | Web search results 31-200 |
| The Malicious Use of Artificial Intelligence: Forecasting, Prevention, and Mitigation | Four High-Level Recommendations | Future of Humanity Institute; University of Oxford; Centre for the Study of Existential Risk; University of Cambridge; Center for a New American Security; Electronic Frontier Foundation; OpenAI | international | Miscellaneous (mixed academic, NPO) | 20-Feb-2018 | unspecified | Citation chaining |
| White Paper: How to Prevent Discriminatory Outcomes in Machine Learning | Executive summary | WEF, Global Future Council on Human Rights 2016-2018 | international | NPO/Charity | 12-Mar-2018 | private sector (companies) | Citation chaining |
| For a meaningful Artificial Intelligence. Towards a French and European strategy | "Part 5 — What are the Ethics of AI?; Part 6 — For Inclusive and Diverse Artificial Intelligence" | Mission Villani | France | Governmental agencies/organizations | 29-Mar-2018 | public sector (French government/parliament) | Linkhubs |
| Statement on Artificial Intelligence, Robotics and 'Autonomous' Systems | Ethical principles and democratic prerequisites | European Commission, European Group on Ethics in Science and New Technologies | EU | IGO/supra-national | xx-Mar-2018 | public sector (EU Commission) | Linkhubs |
| L'intelligenzia artificiale al servizio del cittadino | Sfida 1: Etica | Agenzia per l'Italia Digitale (AGID) | Italy | Governmental agencies/organizations | xx-Mar-2018 | multiple (government, schools, healthcare institutions) | Linkhubs |
| OpenAI Charter | OpenAI Charter | OpenAI | USA | NPO/charity(*) | 9-Apr-2018 | self | Linkhubs |
| AI in the UK: ready, willing and able? (report, only section "An AI Code" taken into account) | no title. P. 125: "… we suggest five overarching principles for an AI Code:" | UK House of Lords, Select Committee on Artificial Intelligence | UK | Governmental agencies/organizations | 16-Apr-2018 | public sector (UK government) | Linkhubs |
| Privacy and Freedom of Expression In the Age of Artificial Intelligence | Conclusions and Recommendations | Privacy International & Article 19 | international | NPO/Charity | 25-Apr-2018 | multiple (states, companies, civil society) | Citation chaining |
| AI Guidelines | AI Guidelines | Deutsche Telekom | Germany | Company | 11-May-2018 | self | Web search results 1-30 |
| The Toronto Declaration: Protecting the right to equality and non-discrimination in machine learning systems | The Toronto Declaration: Protecting the right to equality and non-discrimination in machine learning systems | Access Now ; Amnesty International | international | Miscellaneous (mixed NGO, NPO) | 16-May-2018 | multiple (states, private sector actors) | Linkhubs |
| Discussion Paper on Artificial Intelligence (AI) and Personal Data - Fostering Responsible Development and Adoption of AI | Principles for responsible AI | Personal Data Protection Commission Singapore | Singapore | Governmental agencies/organizations | 5-Jun-2018 | multiple (business; Trade associations and chambers, professional bodies and interest groups) | Linkhubs |
| Our principles | Our principles | Google | USA | Company | 7-Jun-2018 | self | Web search results 1-30 |
| Discussion Paper: National Strategy for Artificial Intelligence (only section "Ethics, Privacy, Security and Artificial Intelligence. Towards a “Responsible AI”") | Ethics, Privacy, Security and Artificial Intelligence. Towards a “Responsible AI” | National Institution for Transforming India (Niti Aayog) | India | Governmental agencies/organizations | 8-Jun-2018 | self (Indian government) | Linkhubs |
| Charlevoix Common Vision for the Future of Artificial Intelligence | Charlevoix Common Vision for the Future of Artificial Intelligence | Leaders of the G7 | international | IGO/supra-national | 9-Jun-2018 | self (gov) | Linkhubs |
| Policy Recommendations on Augmented Intelligence in Health Care H-480.940 | Policy Recommendations on Augmented Intelligence in Health Care H-480.940 | American Medical Association (AMA) | USA | Prof. Association/Society | 14-Jun-2018 | self | Web search results 31-200 |
| Artificial Intelligence: open questions about gender inclusion | Proposals | W20 | international | IGO/supra-national | 2-Jul-2018 | public sector (states/countries) | Web search results 31-200 |
| Everyday Ethics for Artificial Intelligence. A practical guide for designers & developers | Five Areas of Ethical Focus | IBM | USA | Company | 2-Sep-2018 | designers | Web search results 1-30 |
| Artificial Intelligence (AI) in Health | Key recommendations | Royal College of Physicians | UK | Prof. Association/Society | 3-Sep-2018 | multiple (industry, doctors, regulators) | Web search results 31-200 |
| Initial code of conduct for data-driven health and care technology | 10 Principles | UK Department of Health & Social Care | UK | Governmental agencies/organizations | 5-Sep-2018 | developers | Web search results 31-200 |
| Work in the age of artificial intelligence. Four perspectives on the economy, employment, skills and ethics (only section "Good application of artificial intelligence technology and ethics") | Values of a good artificial intelligence society | Ministry of Economic Affairs and Employment | Finland | Governmental agencies/organizations | 10-Sep-2018 | multiple (Finnish world of work) | Linkhubs |
| SAP’s guiding principles for artificial intelligence | SAP’s guiding principles for artificial intelligence | SAP | Germany | Company | 18-Sep-2018 | self | Web search results 1-30 |
| Sony Group AI Ethics Guidelines | Sony Group AI Ethics Guidelines | SONY | Japan | Company | 25-Sep-2018 | self (group) | Web search results 1-30 |
| Ethics Framework - Responsible AI | Framework | Machine Intelligence Garage Ethics Committee | UK | n.a. | 28-Sep-2018 | private sector (start-ups) | Web search results 31-200 |
| Dutch Artificial Intelligence Manifesto | Multidisciplinary challenges | Special Interest Group on Artificial Intelligence (SIGAI), ICT Platform Netherlands (IPN) | Netherlands | Academic and research institution | xx-Sep-2018 | multiple (Dutch government, researchers) | Web search results 31-200 |
| Governing Artificial Intelligence. Upholding Human Rights & Dignity | Recommendations | Data & Society | USA | Research (NPO) | 10-Oct-2018 | multiple (companies, researchers, governments, policy makers, UN) | Citation chaining |
| Tieto's AI ethics guidelines | Tieto's AI ethics guidelines | Tieto | Finland | Company | 17-Oct-2018 | self | Web search results 31-200 |
| Intel's AI Privacy Policy White Paper. Protecting individuals' privacy and data in the artificial intelligence world | Six Policy Recommendations | Intel Corporation | USA | Company | 22-Oct-2018 | public sector (policy makers) | Web search results 31-200 |
| Universal Guidelines for Artificial Intelligence | Universal Guidelines for Artificial Intelligence | The Public Voice | international | Mixed (coalition of NGOs, ICOs etc.) | 23-Oct-2018 | multiple (institutions, governments) | Web search results 1-30 |
| Declaration on ethics and data protection in Artificial Intelligence | "… guiding principles …" | ICDPPC | international | IGO/supra-national | 23-Oct-2018 | unspecified | Web search results 1-30 |
| AI Principles of Telefónica | AI Principles of Telefónica | Telefonica | Spain | Company | 30-Oct-2018 | self | Web search results 1-30 |
| Introducing Unity's Guiding Principles for Ethical AI – Unity Blog | Unity's six guiding AI principles are as follows | Unity Technologies | USA | Company | 28-Nov-2018 | self | Manual inclusion |
| Responsible bots: 10 guidelines for developers of conversational AI | Guideline | Microsoft | USA | Company | xx-Nov-2018 | developers | Manual inclusion |
| AI Now 2018 Report | Recommendations | AI Now Institute | USA | Academic and research institution | xx-Dec-2018 | multiple | Manual inclusion |
| Ethics of AI in Radiology: European and North American Multisociety Statement | Conclusion | American College of Radiology; European Society of Radiology; Radiology Society of North America; Society for Imaging Informatics in Medicine; European Society of Medical Imaging Informatics; Canadian Association of Radiologists; American Association of Physicists in Medicine | international | Prof. Association/Society | 26-Feb-2019 | self | Manual inclusion |
| European ethical Charter on the use of Artificial Intelligence in judicial systems and their environment | "The five principles of the Ethical Charter on the Use of Artificial Intelligence in Judicial Systems and their environment" | Concil of Europe: European Commission for the efficiency of Justice (CEPEJ) | EU | IGO/supra-national | xx-Feb-2019 | multiple (public and private stakeholders) | Manual inclusion |
| Ethically Aligned Design: A Vision for Prioritizing Human Wellbeing with Autonomous and Intelligent Systems, First Edition (EAD1e) | General Principles | Institute of Electrical and Electronics Engineers (IEEE), The IEEE Global Initiative on Ethics of Autonomous and Intelligent Systems | international | Prof. Association/Society | 25-Mar-2019 | multiple (technologists, educators, and policy maker) | Manual inclusion |
| Artificial Intelligence. Australia's Ethics Framework. A discussion Paper | Core principles for AI; A toolkit for ethical AI | Department of Industry Innovation and Science | Australia | Governmental agencies/organizations | 5-Apr-2019 | unspecified | Manual inclusion |
| Ethics Guidelines for Trustworthy AI | Ethical Principles in the Context of AI Systems | High-Level Expert Group on Artificial Intelligence | EU | IGO/supra-national | 8-Apr-2019 | multiple (all stakeholders) | Manual inclusion |
| Ethical, social, and political challenges of Artificial Intelligence in Health | Conclusion | Future Advocacy | UK | Company | xx-Apr-2019 | unspecified | Manual inclusion |
| The responsible AI framework | Operating AI | PriceWaterhouseCoopers UK | UK | Company | n.a. | multiple (clients) | Web search results 31-200 |
| Digital Decisions | VI. Solutions Part 1: Principles | Center for Democracy & Technology | USA | NPO/charity | n.a. | unspecified | Citation chaining |
| Responsible AI and robotics. An ethical framework. | Our view | Accenture UK | UK | Company | n.a. | private sector | Web search results 1-30 |
| Commitments and principles | OP Financial Group's ethical guidelines for artificial intelligence | OP Group | Finland | Company | n.a. | self | Web search results 31-200 |
| Science, Law and Society (SLS) Initiative | Principles for the Governance of AI | The Future Society | USA | NPO/charity | n.a. | public sector (policy makers) | Linkhubs |
| 10 Principles of responsible AI | Summary of our proposed Recommendations | Women leading in AI | n.a. | n.a. | n.a. | public sector (national and international policy makers) | Manual inclusion |
| AI4People—An Ethical Framework for a Good AI Society: Opportunities, Risks, Principles, and Recommendations | Action Points | AI4People | EU | n.a. | n.a. | unspecified | Manual inclusion |
| AI Principles & Ethics | AI Principles; AI guidelines | Smart Dubai | UAE | Governmental agencies/organizations | n.a. 2018? | self | Manual inclusion |

## Table S2. Screening and Eligibility (details)

| | |
|---|---|
| **Screening** | |
| Sources considered: | - Types: websites and documents published online or parts thereof such as policy documents, principles, guidelines, recommendations, dedicated webpages, institutional reports and declarations; |
| | - Issuers: institutions, associations and organizations such as companies, corporations, NGOs, NPOs, academic and professional societies, governmental institutions and affiliated organizations; |
| | - Language: English, German, French, Italian, Greek (the languages spoken by the researchers). |
| Sources excluded: | - Types: videos, images, audio/podcasts, books, blog articles, academic articles, journalistic articles, syllabi, legislation, official standards, conference summaries; |
| | - Issuers: individual authors; |
| | - Language: others than those above. |
| **Eligibility** | |
| Sources included: | - which refer to "artificial intelligence" and/or "AI", either explicitly in their title or within their description (example: UK, House of Lords: "AI in the UK: ready, willing and able"); or |
| | - which do not contain the above reference in their title but mention "robot" or "robotics" instead *and* reference AI or artificial intelligence explicitly as being part of robots and/or robotics (example: "Principles of robotics"); or |
| | - which do not contain the above reference in their title but are thematically equivalent (by referring to "algorithms", "predictive analytics", "cognitive computing", "machine learning", "deep learning", "autonomous" or "automated" instead (example: "Automated and Connected Driving: Report"). |
| | - which self-proclaim to be a principle or guideline (including "ethics/ethical", "principles", "tenets", "declaration", "policy", "guidelines", "values" etc.); or |
| | - which is expressed in normative or prescriptive language (i.e. with modal verbs or imperatives); or |
| | - which is principle- or value-based (i.e. indicating a preference and/or a commitment to a certain ethical vision or course of action). |
| Excluded sources: | - websites and documents about robotics that do not mention artificial intelligence as being part of robots/robotics; and |
| | - websites and documents about data or data ethics that do not mention artificial intelligence as being part of data; |

## Table S3. Categorization after themeing and code mapping

| Question addressed | Thematic family | Themes |
|---|---|---|
| **What?** | **Ethical Principles & Values** | **Ethical Principles**<br>I. Beneficence<br>II. Non-maleficence<br>III. Trust<br>IV. Transparency & Explainability<br>V. Freedom and autonomy (incl. consent)<br>VI. Privacy<br>VII. Justice, Fairness & Equity<br>VIII. Responsibility & Accountability<br>IX. Dignity<br>X. Sustainability<br>XI. Solidarity |
| | **Technical and methodological aspects** | **Specific functionalities**<br>I. Feedback & feedback-loop<br>II. Decision-making |
| | | **Data & datasets**<br>I. Data origin/input<br>II. Data use<br>III. Metadata<br>IV. Algorithms |
| | | **Methodological challenges**<br>I. Methodology<br>II. Metris & measurements<br>III. Tests, testing<br>IV. Ambiguity & uncertainty<br>V. Accuracy<br>VI. Reliability<br>VII. Evidence and validation<br>VIII. Black-box (opacity)<br>IX. Data security<br>X. Quality (of data/system/etc.) |
| | **Impact** | **Benefits**<br>I. AI strengths, advantages<br>II. Knowledge<br>III. Innovation<br>IV. Enhancement |
| | | **Risks**<br>I. Risks<br>II. Malfunction<br>III. Misuse & dual-use<br>IV. Deception<br>V. Discrimination (duplicate in Justice&Fairness)<br>VI. Surveillance<br>VII. Manipulation<br>VIII. Arms race |
| | | **Impact assessment**<br>I. Impact<br>II. Goals/Purposes/Intentions<br>III. Public opinion<br>IV. Risk evaluation & mitigation (duplicate in Risks)<br>V. Monitoring/Precaution<br>VI. Future of work |
| **Who?** | **Design & development** | I. Industry<br>II. AI researchers<br>III. Designers<br>IV. Developers |
| | **Users** | I. End users<br>II. Organisations<br>III. Public sector actors<br>IV. Military<br>V. Communities |
| | **Specific stakeholders** | I. Ethical and/or auditing committees<br>II. Government<br>III. Policy makers<br>IV. Researchers & scientists<br>V. Vulnerable groups & minorities |
| **How?** | **Social engagement** | I. Knowledge commons<br>II. Education & training<br>III. Public deliberation & democratic processes<br>IV. Stakeholder involvement & partnerships |
| | **Soft policy** | I. Standards<br>II. Certification<br>III. Best practices<br>IV. Whistleblowing |
| | **Economic incentives** | V. Business model & strategy<br>VI. Funding & investments<br>VII. Taxes/taxation |
| | **Regulation & audits** | VIII. Laws & regulation (general)<br>IX. Data protection regulation<br>X. IP law<br>XI. Human rights treaties<br>XII. Other rights & laws<br>XIII. Audits & auditing |